\documentclass[final]{IEEEtran}
\usepackage{algorithm}
\usepackage{color}
\usepackage{algpseudocode}
\usepackage{graphics}
\usepackage{epsfig}
\usepackage{cite}
\usepackage{threeparttable}
\usepackage{graphicx}
\usepackage{picinpar}
\usepackage{subfigure}
\usepackage{amssymb}
\usepackage[cmex10]{amsmath}
\usepackage{url}

\usepackage{bm}
\usepackage{setspace}
\usepackage{enumerate}
\usepackage{makecell}

\usepackage{multirow}
\usepackage{array}
\usepackage{amssymb}

\input epsf

\newcommand*{\myLargeBullet}{\vcenter{\hbox{\Large$\bullet$}}}
\begin{document}

\title{\huge{Intelligent Metasurface-Enabled Integrated Sensing and Communication: Unified Framework and Key Technologies
} }

\author{Shunyu Li,  
Tianqi Mao, \IEEEmembership{Member,~IEEE},
Guangyao Liu, 
Fan Zhang, 
Ruiqi Liu, \IEEEmembership{Senior Member,~IEEE},

Meng Hua, \IEEEmembership{Senior Member,~IEEE},
Zhen Gao, \IEEEmembership{Member,~IEEE},
Qingqing Wu, \IEEEmembership{Senior Member,~IEEE},

and George K. Karagiannidis,~\IEEEmembership{Fellow,~IEEE}


\thanks{S. Li and Z. Gao are with the State Key Laboratory of CNS/ATM, Beijing Institute of Technology, Beijing, China. (e-mails: li.shunyu@bit.edu.cn, gaozhen16@bit.edu.cn).}
\thanks{ T. Mao (\textit{Corresponding Author}) is with Greater Bay Area Innovation Research Institute of BIT, Zhuhai, China, and is also with the State Key Laboratory of Environment Characteristics and Effects for Near-space, Beijing Institute of Technology, Beijing, China (e-mail: maotq@bit.edu.cn).}
\thanks{G. Liu is with Beihang University, Beijing 100191, China (e-mail: liugy@buaa.edu.cn).}
\thanks{F.~Zhang is with Tsinghua University, Beijing 100084, China (e-mail: zf22@mails.tsinghua.edu.cn).}
\thanks{R. Liu is with ZTE Corporation, Beijing 100029, China, (e-mail: richie.leo@zte.com.cn).}
\thanks{M. Hua is with Imperial College London, SW7 2AZ London, U.K. (e-mail: m.hua@imperial.ac.uk).}
\thanks{Q. Wu is with Shanghai Jiao Tong University, Shanghai 200240, China, (e-mail: qingqingwu@sjtu.edu.cn).}
\thanks{G. K. Karagiannidis is with Aristotle University of Thessaloniki,54124 Thessaloniki, Greece (e-mail: geokarag@auth.gr).}
\vspace{-3mm}} %

\maketitle
\begin{abstract}
As the demand for ubiquitous connectivity and high-precision environmental awareness grows, integrated sensing and communication (ISAC) has emerged as a key technology for sixth-generation (6G) wireless networks.
Intelligent metasurfaces (IMs) have also been widely adopted in ISAC scenarios due to their efficient, programmable control over electromagnetic waves. This provides a versatile solution that meets the dual-function requirements of next-generation networks.
Although reconfigurable intelligent surfaces (RISs) have been extensively studied for manipulating the propagation channel between base and mobile stations, the full potential of IMs in ISAC transceiver design remains under-explored.
Against this backdrop, this article explores emerging IM-enabled transceiver designs for ISAC systems. It begins with an overview of representative IM architectures, their unique principles, and their inherent advantages in EM wave manipulation.
Next, a unified ISAC framework is established to systematically model the design and derivation of diverse IM-enabled transceiver structures. This lays the foundation for performance optimization, trade-offs, and analysis.
The paper then discusses several critical technologies for IM-enabled ISAC transceivers, including dedicated channel modeling, effective channel estimation, tailored beamforming strategies, and dual-functional waveform design.

\end{abstract}

\vspace{-5mm}
\section{Introduction}
The evolution toward sixth-generation (6G) network brings a fundamental rethinking of the relationship between electromagnetic (EM) signal transmission and the surrounding EM environment, where integrated sensing and communication (ISAC) has emerged as one promising philosophy that enables both functionalities to coexist with superior spectral, energy and cost efficiencies \cite{MAO_TCOM,FANZ_JSAC}. For instance, the ISAC technology can support ubiquitous sensing with existing networking infrastructures, which enables various cutting-edge applications like autonomous driving and digital twin.

The multi-input multi-output (MIMO) architecture has been extensively explored in ISAC to exploit spatial degrees of freedom (DoFs) \cite{MIMO_ISAC}.
However, the ever-increasing dimensions of the antenna array introduce non-trivial challenges for MIMO-ISAC systems.
In particular, the deployment of numerous radio-frequency (RF) chains and complex feeding networks, comprising power-hungry and costly components such as amplifiers, phase shifters, and microstrip lines, can result in excessive power consumption and hardware overhead. 
These limitations significantly hinder the practical implementation of MIMO-ISAC, particularly in scenarios constrained by energy or budget \cite{GAO_MIMO}.

In this context, intelligent metasurface (IM)-enabled ISAC transceiver design has surfaced as a compelling alternative, where various forms of IMs, such as reconfigurable intelligent surfaces (RIS), stacked intelligent metasurfaces (SIM), dynamic metasurface antennas (DMA), and reconfigurable holographic surfaces (RHS), can be deployed at the base station (BS) to support ISAC functionalities \cite{TRIS_2,TRIS_1,SIM_me,SIM_2,DMA_1,DMA_2,RHS_1}. 
Tailored IM-enabled transceiver design can directly manipulate the EM properties of its meta-elements, allowing it to achieve equivalent precoding/combining functions to traditional phased-array counterparts.
This structurally obviates the necessity for massive RF chains and the intricate feeding network in classical fully-digital/hybrid beamformers, leading to more compact, economical, and energy-efficient ISAC systems \cite{SIM_me}.
Additionally, thanks to the recent progress in metamaterial fabrication, the size and adjacent spacing of meta-elements in IM-enabled antenna arrays can be accomplished on a sub-wavelength scale, overcoming the half-wavelength restriction inherent in conventional array-based architectures.  
This allows IM-enabled transceivers to accommodate substantially more antenna elements within a given aperture size. 
As element spacing decreases, extremely large-scale antenna array (XLAA) technology becomes feasible, supporting holographic MIMO capabilities through quasi-continuous control over EM field distributions for superior ISAC performance \cite{DMA_2}.

\begin{figure*}[t]
    \centering
    \includegraphics[width=0.8\linewidth]{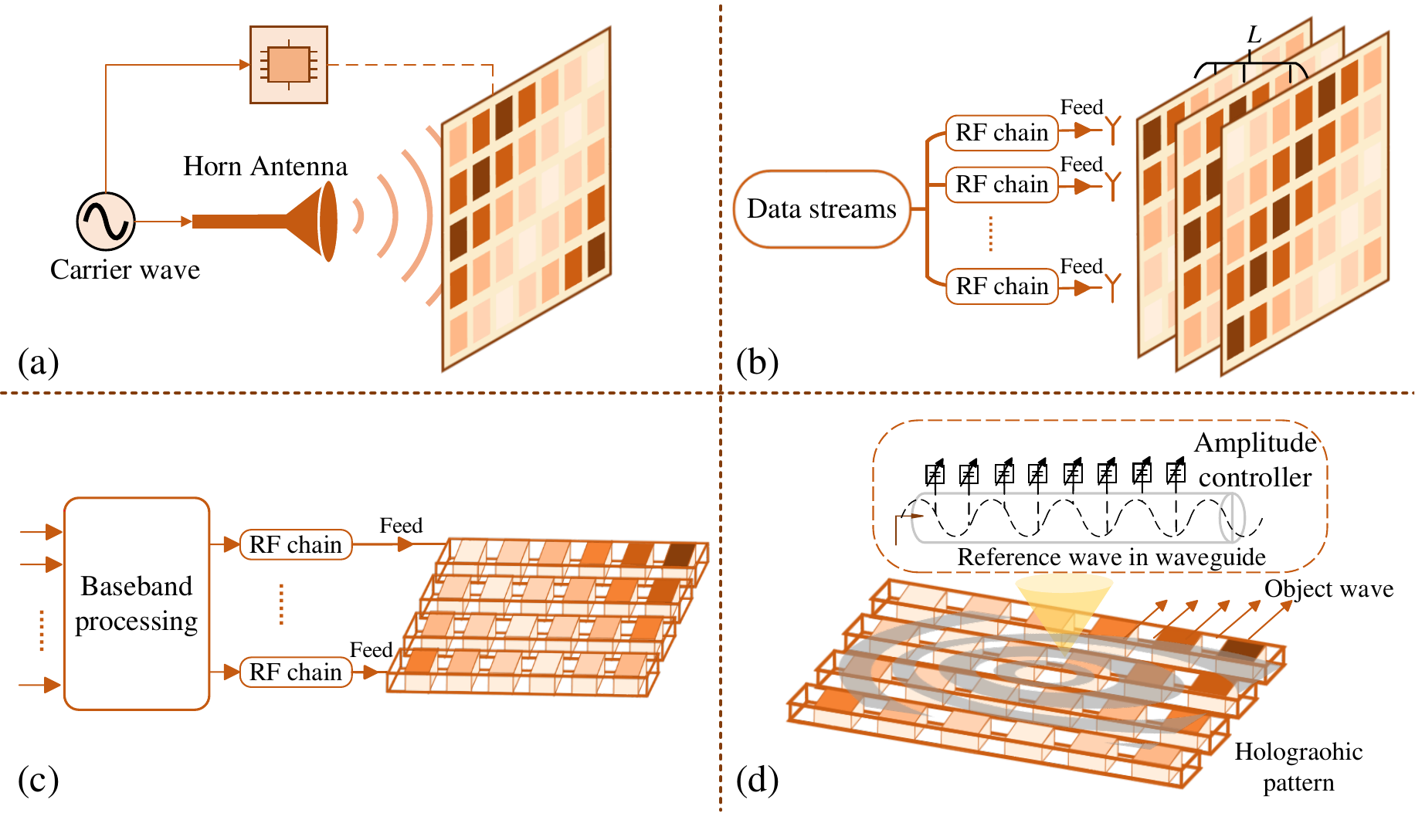}
    \caption{Illustration of existing IM-enabled ISAC architectures: 
a) RIS \cite{TRIS_1};
b) SIM \cite{SIM_2};
c) DMA \cite{DMA_1}; 
d) RHS \cite{RHS_1}. }
    \label{fig:1}
    \vspace{-3mm}
\end{figure*}

Against this background, this article presents a comprehensive overview of IM-enabled transceiver design for ISAC systems, hereafter referred to as \emph{IM-enabled ISAC}. 
Specifically, we first introduce their fundamental characteristics and architectural diversity of various IM designs.
These architectures are then unified under a common ISAC framework, facilitating consistent analysis and implementation across different configurations.
Following that, we explore the key technologies of IM-enabled ISAC systems, spanning channel modeling, channel estimation, tailored beamforming strategies, and integrated waveform design, and discuss their corresponding solutions.


\section{Unified Framework for IM-enabled ISAC}
In this section, we present a comparative analysis of emerging IM-enabled transceiver architectures and discuss how their structural differences lead to unique mechanisms for signal generation and manipulation in ISAC applications.
Afterward, a general framework is proposed to effectively characterize these diverse implementations, offering a foundation for systematic investigation and theoretical modeling of IM-enabled ISAC systems.
\begin{table*}[!t]
{
\caption{Comparison of Different IM-enabled ISAC Transceiver Architectures}
}
\label{tab:compare}
\centering
\begin{tabular}{|c|c|c|c|c|}
\hline
\textbf{Feature} & \textbf{DMA} & \textbf{RHS} & \textbf{RIS} & \textbf{SIM} \\
\hline
Structure complexity & Medium & Medium & Simple & Complex \\
\hline
Propagation mode & Serial & Serial & Parallel & Parallel \\
\hline
Baseband digital precoding & Yes & Yes & No & {Optional (typically no)} \\
\hline
RF chain count & Multiple & Multiple & Single & Multiple \\
\hline
Signal from feed source & {Information-bearing} & {Information-bearing} & Carrier only & {Information-bearing} \\
\hline
Feed source position & Embedded & Embedded & External & External \\
\hline
Element reconfiguration & {Coupled amplitude-phase} & Amplitude only & {Phase only} & {Phase only} \\
\hline
Element constraint & {Lorentzian constraint} & Amplitude discrete & Unit-modulus & Unit-modulus \\
\hline
Hardware efficiency & Medium & High & Very high & Medium \\
\hline
Information modulation & At RF chains & At RF chains & At metasurface & At RF chains \\
\hline
Waveguide required & Yes & Yes & No & No \\
\hline
Spatial DoF & Medium & Medium & Limited & High \\
\hline
Power efficiency & Medium & Medium & High & Medium-high \\
\hline

\end{tabular}
\end{table*}

\subsection{Existing IM-enabled ISAC Transceiver Design}

{\bf Reconfigurable Intelligent Surface}:
As one of the most structurally simple and cost-effective IM architecture, RIS transceiver typically consists of a horn antenna and a planar surface of numerous low-power, passive reconfigurable meta-elements, as shown in Fig.~\ref{fig:1}a.
Here, the feed antenna only emits a carrier wave, while the modulation of communication data streams or specific waveform shaping for sensing is achieved by dynamically adjusting the EM responses of the RIS elements \cite{TRIS_2}.
This approach effectively eliminates the need for digital precoding and only requires one RF chain to realize the desired dual-functions, resulting in lower power consumption and cost.
Specifically, by varying their EM responses over time, the RIS elements can process the incident carrier wave to generate multiple frequency harmonics. These harmonics can then be independently shaped and spatially steered into distinct beams, enabling dual-function operation with minimal complexity \cite{SPACE_TIME}. 

{\bf Stacked Intelligent Metasurface}:
SIM utilizes multiple cascaded transmissive metasurface layers that facilitate sophisticated wave-domain signal processing through their enhanced EM wave manipulation capabilities.
As shown in Fig.~\ref{fig:1}b, SIM mainly comprises multiple parallel metasurfaces with sub-wavelength separation, fed by external RF chains \cite{SIM_2}.
When signals from the RF chains propagate through these metasurfaces, a cascaded coupling effect is established. This effect begins with the initial signals from the RF chains fully engaging all elements on the first layer. 
Subsequently, elements on each metasurface re-radiate EM waves, which in turn serve as the input illuminating all elements of the next metasurface in the cascade. 
Wave-domain signal processing thus occurs sequentially as these waves traverse the entire multi-layer structure, with phase shifts at each meta-element on each layer reconfiguring the wavefront, akin to a physical neural network.
This multi-layer structure offers vastly increased spatial DoF, enabling the generation of highly complex and tailored beam patterns for enhanced communication and sensing performance \cite{SIM_me}.  
On the other hand, the intricate interlayer coupling inherent in SIM presents considerable challenges for ISAC beamforming, necessitating meticulous optimization of all elemental phase shifts to achieve desired dual-functionality, which will be elaborated on in Section~\ref{sec:key}.
{The SIM architecture considered in this work follows the widely adopted baseline implementation without baseband digital precoding. Nevertheless, an optional variant incorporating a digital precoder upstream of the RF chains can also be envisioned, whose modeling is accommodated within the proposed unified framework and elaborated in the subsequent subsection.}

{\bf Dynamic Metasurface Antenna}: 
DMA is an emerging concept within the paradigm of ``radiative metasurfaces'', which uniquely performs sophisticated analog signal processing directly within a waveguide-integrated radiating structure \cite{DMA_1}.
As shown in Fig.~\ref{fig:1}c, the DMA architecture consists of multiple configurable meta-elements situated on a set of one-dimensional waveguides (typically implemented as microstrip lines).
Each RF chain is connected to  a distinct waveguide and exclusively controls the meta-elements situated on that path, forming a partially connected hybrid structure similar to conventional array antennas {\cite{DMA_2}}.
Under this distinctive architecture, the independent signal streams from RF chains are fed in parallel to their respective waveguides.
Each stream sequentially traverses a series of meta-elements, where each element reconfigures and radiates the incoming signal into the wireless channel based on its frequency response.
The signal manipulation in the DMA is predominantly influenced by two factors: the propagation characteristics within the waveguide and the reconfigurable response of the radiating elements \cite{DMA_1}.
The former determines the propagation attenuation and phase shifts, which originate from the dielectric loss and conductor loss that depend on the material properties of the microstrip line and carrier frequency.
The latter pertains to {the meta-elements’ resonant Lorentzian frequency response. Under a normalized narrowband surrogate around the operating frequency, it reduces to a frequency-independent complex-plane parameterization where the radiated amplitude is intrinsically coupled to the desired phase shift \cite{DMA_1}.}
Nevertheless, DMA can synthesize highly complex radiation patterns through precise, collective control of these sub-wavelength spacing tunable elements, enabling advanced generation of tailored concurrent communication and sensing beams.

{\bf Reconfigurable Holographic Surface}: 
RHS is an IM-enabled transceiver that incorporates the philosophies from leaky-wave and holographic antennas \cite{RHS_1}.
{Structurally, an RHS shares similarities with a DMA: as illustrated in Fig.~\ref{fig:1}d, it comprises a parallel-plate waveguide where RF signals are fed and propagate sequentially past each radiation element. 
The key distinction lies in the function of the sub-wavelength meta-elements: instead of resonance tuning as in DMA, RHS primarily controls the local leakage rate (i.e., radiated amplitude) along the waveguide.}
Unlike Holographic MIMO (HMIMO), which typically involves (quasi-)continuous apertures (e.g., as in \cite{FPWE}) and is further detailed in Section~\ref{sec:channel}, the ``holographic'' aspect of RHS is rooted in optical holographic principles for wavefront reconstruction.
In this holographic paradigm, the guided wave traveling along the waveguide serves as the reference wave, while the desired beam pattern or EM field distribution acts as the object wave. 
The configurable amplitude profile of the RHS elements then functions as a holographic pattern, which records the interference produced by the coherent superposition between these two waves. 
When illuminated by the reference wave, this holographic pattern reconstructs the corresponding object wave, thereby achieving the required precoding/combining functionality.
Therefore, by modulating the voltage applied to each element to meticulously regulate its radiation amplitude, RHS can dynamically reconfigure its holographic pattern. This allows it to generate required object waves (beams) for both communication and sensing.
Compared to traditional array-based architecture, RHS can provide lower power consumption and hardware cost by eliminating the need for phase-shifting circuit \cite{RHS_1}.

A summary comparison of the aforementioned IM-enabled ISAC transceiver designs is presented in Table~\ref{tab:compare}. These distinct physical configurations lead to unique wave manipulation mechanisms and signal processing principles. Despite their technological diversity, a systematic analytical framework is necessary to provide a unified perspective on how these distinct designs function in ISAC scenarios, establishing common theoretical foundations applicable to both existing architectures and potential future variants.

\subsection{Unified Framework for IM-enabled ISAC Systems}
\begin{figure*}[t]
    \centering
    \includegraphics[width=1\linewidth]{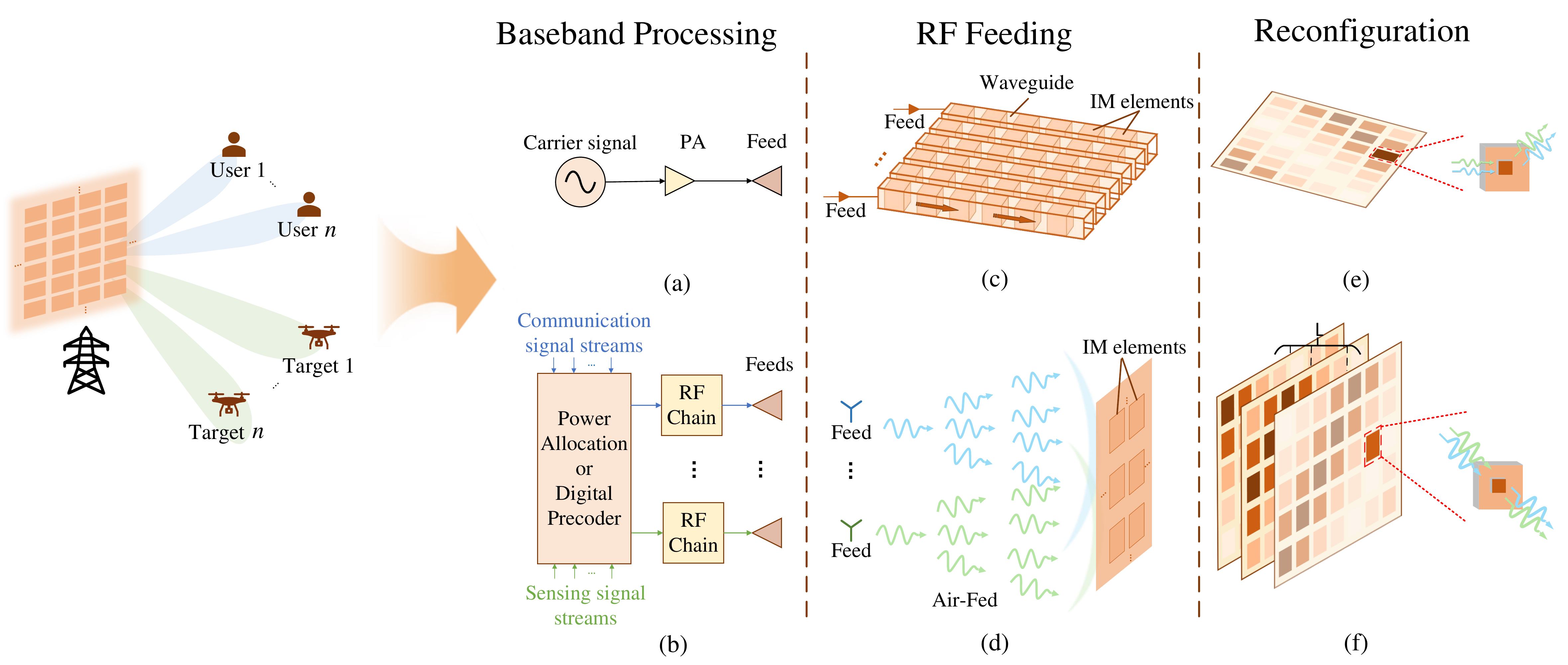}
    \caption{Illustration of the unified framework for IM-enabled ISAC systems.}
    \label{fig:2}
    \vspace{-1mm}
\end{figure*}

This article considers a monostatic scenario, where an IM-enabled BS is deployed for simultaneous communication and sensing with multi-user/target beamforming, as illustrated in Fig.~\ref{fig:2}. 
To establish a general analytical foundation for the diverse IM-enabled ISAC architectures exemplified above, we propose a unified framework that delineates the entire signal generation and processing mechanism for both communication and sensing functionalities, where the transmitter model is emphasized.

\begingroup

\begin{table*}[!t]
\centering
\begin{threeparttable}
\caption{Manifestations of Unified Framework Components \\
\small (Markers: $\myLargeBullet$ RIS, $\bigstar$ SIM, $\blacksquare$ DMA, $\blacktriangle$ RHS)}
\label{tab:framework}
\begin{tabular}{|>{\centering\arraybackslash}m{4cm}|>{\centering\arraybackslash}m{4.6cm}|>{\centering\arraybackslash}m{4.5cm}|>{\centering\arraybackslash}m{2.1cm}|}
\hline
\makecell[c]{\textbf{Component}} & \makecell[c]{\textbf{Primary Distinguishing}\\\textbf{Feature}} & \makecell[c]{\textbf{Specific Realization /}\\\textbf{Constraint}} & \makecell[c]{\textbf{Example}\\\textbf{Architectures}} \\
\hline

\multirow{3}{*}{\parbox{4cm}{\centering Baseband Processing Matrix, $\mathbf{V}$}}
 & \makecell[c]{Single RF Chain} & \makecell[c]{Power of the Carrier Signal} & \makecell[c]{$\myLargeBullet$} \\ 
\cline{2-4}
 & \multirow{2}{*}{\makecell[c]{Multiple RF Chains}} & \makecell[c]{Digital Precoding Matrix} & \makecell[c]{$\blacksquare$ $\blacktriangle$ {$(\bigstar)$\tnote{*}}} \\
\cline{3-4}
 & & \makecell[c]{Power Allocation Vector} & \makecell[c]{$\bigstar$} \\
\hline

\multirow{2}{*}{\parbox{4cm}{\centering RF Feeding Matrix, $\mathbf{T}_l$}}
 & \makecell[c]{Waveguide (Serial)} & \makecell[c]{Block-Diagonal Matrix} & \makecell[c]{$\blacksquare$ $\blacktriangle$} \\
\cline{2-4}
 & \makecell[c]{Free-Space (Parallel/Inter-layer)} & \makecell[c]{Dense Channel Matrix} & \makecell[c]{$\myLargeBullet$ $\bigstar$} \\ 
\hline

\multirow{3}{*}{\parbox{4cm}{\centering Reconfiguration Matrix, $\mathbf{Q}_l$}}
 & \makecell[c]{Amplitude-Only} & \makecell[c]{Amplitude Range} & \makecell[c]{$\blacktriangle$} \\
\cline{2-4}
 & \makecell[c]{Phase-Only} & \makecell[c]{Unit-Modulus} & \makecell[c]{$\myLargeBullet$ $\bigstar$} \\ 
\cline{2-4}
 & \makecell[c]{Coupled Amp \& Phase} & \makecell[c]{Lorentzian Constraint} & \makecell[c]{$\blacksquare$} \\
\hline

\end{tabular}
{
\begin{tablenotes}

    \item[*] \small{Represents an optional variant configuration for the architecture.}
\end{tablenotes}
}
\end{threeparttable}
\end{table*}

\endgroup
As shown in Fig.~\ref{fig:2}, the proposed framework conceptually disaggregates the transmit behavior into sequential functional slices, providing a composable methodology to model/analyze different implementations through appropriate selection of slicing components (as illustrated in Table~\ref{tab:framework}).
More specifically, the proposed unified framework includes three essential processes that can characterize the entire transmit behavior of different IM-enabled ISAC design: {\bf baseband processing,  RF feeding}, and {\bf reconfiguration}. 
These processes collectively offer a clear and structured interpretation of how signals are generated, radiated, and beamformed at the transmitter side, as detailed in the following discussion.

As illustrated in Figs.~\ref{fig:2}a and 2b, the first slice in our unified framework is the {\bf baseband processing}, where the {\bf Baseband Processing Matrix ($\mathbf{V}$)} processes the input data streams destined for the RF chains.
Within our unified framework, this process is categorized into three primary types according to the presence of digital precoding and the number of RF chains in the IM-enabled ISAC system, as clarified in Table~\ref{tab:framework}.
For architectures equipped with a digital precoder, such as DMA and RHS, the baseband processing maps multiple data streams to the corresponding RF chains to enable spatial multiplexing. 
For other architectures that forgo digital precoding, the baseband process is concerned with power management. 
{
In a single-RF-chain configuration such as RIS, baseband processing reduces to simply assigning a scalar power level to the carrier signal.
For the optional (typically not adopted) SIM variant with baseband digital precoding, the baseband processing becomes a digital precoding matrix applied upstream of the RF chains.}

The second component, {\bf RF feeding}, characterized by the {\bf RF Feeding Matrix ($\mathbf{T}_l$)}, serves as the crucial slice between the RF chains and the meta-elements, where the subscript corresponds to the $l$-th layer.
In multilayer architectures, such as SIM, the transmission from the initial sources to the first layer is represented by $\mathbf{T}_1$, while the subsequent inter-layer transmission from a preceding layer to the $l$-th layer is naturally characterized by its own feed matrix, $\mathbf{T}_l$. 
As summarized in Table~\ref{tab:framework}, its mathematical form depends on the specific architecture. 
For systems with embedded waveguides like DMA and RHS (see Fig.~\ref{fig:2}c), their inherent partially connected topology dictates a block-diagonal structure for the RF feeding Matrix. Each block in this matrix encapsulates the cumulative amplitude attenuation and phase shifts along an independent waveguide path.
In contrast, for externally-fed systems such as RIS and SIM (see Fig.~\ref{fig:2}d), the parallel feeding mechanism creates a fully-connected topology, which is captured by a dense RF feeding Matrix. 
For a SIM with multiple RF chains, this matrix describes the mapping from all sources to all meta-elements, which is typically a diffraction matrix derived from physical optics, characterizing the complex wave travel between layers \cite{SIM_me,SIM_2}. 
Furthermore, the RF feeding matrix for RIS is significantly simplified; given that it typically employs only a single RF chain  and its feed emits an unmodulated carrier wave, the entire RF feeding link is commonly reduced to a single constant factor for analytical tractability, which captures the overall power attenuation \cite{TRIS_1}.

The final slice in our framework is the {\bf reconfiguration } process, where the metasurfaces alter the wavefront by applying element-wise phase and/or amplitude modifications (see Figs.~\ref{fig:2}e and 2f). 
This is characterized by the {\bf Reconfiguration Matrix ($\mathbf{Q}_l$)}.
In multi-layer architectures like SIM, the signal processing takes on a cascaded form, where the RF feeding matrix $\mathbf{T}_l$ and the reconfiguration matrix $\mathbf{Q}_l$ for each layer act as a pair, representing the sequential processes of signal transmission to, and reconfiguration at, the $l$-th layer.
When inter-element coupling effects are neglected, the reconfiguration matrix is typically diagonal. Each diagonal entry thus represents a specific complex coefficient applied by a single meta-element to its local incident signal. 
The ability to adjust these complex coefficients is  the key enabler of reconfigurability in an IM-enabled ISAC system. 
As summarized in Table~\ref{tab:framework}, this tuning capability manifests as distinct EM control modalities, each regulated by a specific  constraint. 
These modalities include amplitude-only control within a {prescribed range} (e.g., RHS), phase-only control {under} a unit-modulus constraint (e.g., RIS and SIM), and {coupled amplitude–phase control } governed by a Lorentzian constraint (e.g., DMA).

However, it is worth noting that modeling the reconfiguration matrix as diagonal is merely an approximation.
For a high-fidelity analysis of arrays with subwavelength spacing where interelement coupling is dominant, the reconfiguration effect is collective rather than element-wise.
This requires a non-diagonal reconfiguration matrix.
A rigorous method for determining this matrix's entries involves multiport network analysis, in which an impedance matrix formalism is often used to characterize complete voltage-current relationships across all element terminals. This formalism inherently accounts for all cross-element interactions.\cite{multiport}.
Furthermore, when considering the transition to a quasi-continuous aperture, this discrete impedance matrix formalism evolves into a continuous surface current distribution. Correspondingly, the channel model must also be adapted to this continuous domain, a topic detailed in Section.~\ref{sec:channel}.

With the proposed unified framework, various IM-enabled ISAC systems can be effectively characterized.
On one hand, the received communication signal at users can be expressed~as
\begin{equation}
\label{eq:1}
\mathbf{Y}=\mathbf{H}\left(\prod_{l=1}^L\mathbf{Q}_l\mathbf{T}_l\right)\mathbf{V}\mathbf{x}+\mathbf{n},
\end{equation}
where the baseband processing matrix $\mathbf{V}$, RF feeding matrices $\mathbf{T}_l$ and reconfiguration matrices $\mathbf{Q}_l$ jointly determine the RF signal transmitted by $L$-layer IM-enabled antennas ($L\geq 1$). 
Here, $\mathbf{x}$ represents the source signal vector prior to baseband processing.
Besides, $\mathbf{H}$ and $\mathbf{n}$ denote the  wireless channel and additive white Gaussian noise, respectively. 
This formulation serves as a foundation for communication performance optimization, such as classical sum-rate maximization.

On the other hand, for multi-beam sensing, the beam pattern towards critical targets is usually considered, which can be formulated~as

\begin{equation}
\label{eq:2}
\mathbf{P}=\mathbf{a}^H\left(\prod_{l=1}^L\mathbf{Q}_l\mathbf{T}_l\right)\mathbf{V}\mathbf{V}^H\left(\prod_{l=1}^L\mathbf{T}_l^H\mathbf{Q}_l^H\right)\mathbf{a},
\end{equation}
where $\mathbf{a}$ represents the steering vector of the last-layer IM towards the target direction. This expression provides a unified representation of the angular-domain energy distribution of the sensing beams, which is applicable to different IM structures regarding beam pattern optimization.

{
Moreover, the framework can be extended to time‑varying IMs by modeling the reconfiguration as a time‑periodic operator and expanding it into Fourier harmonics, after which \eqref{eq:1}-\eqref{eq:2} are applied to each harmonic without changing their form. This provides a unified description of spatial beamforming and spectral shaping at each generated frequency, consistent with established practices in time-modulated arrays \cite{TRIS_1, SPACE_TIME}.
}


In conclusion, through appropriate modular matrix representations, this unified framework accommodates diverse architectures, from single-layer RIS to multi-layer SIM configurations, while featuring various feeding mechanisms and optional digital precoding procedures. Beyond capturing existing designs, the proposed framework can also characterize some variant configurations, such as the digitally precoded SIM systems.
Such a framework will definitely benefit systematic understanding of the IM-enabled ISAC, providing useful insights  especially for its performance analysis and dual-functional optimization.

\section{Key Technologies for IM-enabled ISAC Systems }

Building upon the proposed unified framework, IM-enabled ISAC systems exhibit distinctive  signal interaction mechanisms due to their distinctive structure, necessitating specialized approaches to achieve desired dual-functional performance.
In this section, we explore key technologies for IM-enabled ISAC systems, focusing on dedicated channel modeling, channel estimation, tailored beamforming strategies, and dual-functional waveform design.

\subsection{Channel Modeling for IM-enabled ISAC}
\label{sec:channel}
The large apertures enabled by IMs, foundational to XLAA, significantly extend the Rayleigh distance. 
As a result, radiative near-field or mixed near-far-field propagation conditions become prevalent in typical ISAC scenarios.
In these regions, the spherical wave propagation model becomes essential. The resulting channel steering vectors acquire a non-linear phase component explicitly tied to element-to-target/user distances. This distance-resolvable phase information creates a paradigm shift for ISAC. For communication, it allows distinguishing users at the same angle but different distances; for sensing, it imbues received signals with richer information, leading to more precise range and velocity estimations.

The sub-wavelength structure of IMs enables effectively continuous apertures, a concept central to HMIMO that demands a fundamental, physics-based channel model. While the channel behavior is governed by the electromagnetic Green's function, its continuous spatial-domain form is computationally intractable. Fourier plane-wave series expansion  elegantly solves this by transforming the spatial Green's function into the wavenumber domain. 
This method decomposes the complex wave propagation into a discrete superposition of orthogonal plane waves \cite{FPWE}. Critically, this reveals that the channel's intrinsic DoFs are determined by the array's physical aperture and the scattering environment, not simply by the number of elements. 
Ultimately, this high-fidelity channel model lays the theoretical foundation for wavefield synthesis. It supports the ability to perform high-resolution environmental fingerprinting for sensing and communication, to achieve a massive leap in performance by approaching the fundamental capacity of holographic channels.

\subsection{Channel Estimation of IM-enabled ISAC}
To enable the practical implementation of the subsequent integrated beamforming and waveform design, acquiring precise channel state information (CSI) is a critical prerequisite. 
A primary challenge lies in the massive number of meta-elements, which leads to a high-dimensional estimation problem and imposes prohibitive pilot overhead when using conventional methods.
Furthermore, the passive nature of meta-elements requires CSI to be estimated indirectly. In architectures like DMA and SIM, the RF chains are far outnumbered by the meta-elements. Consequently, the full CSI must be inferred from the signals at these few RF chains, each capturing a superposition of responses from numerous elements, a process dictated by the RF feeding slice and necessitating precise accounting during estimation.

To overcome these challenges, novel estimation methods are being developed. 
For SIM, where the estimation is an underdetermined problem, one advanced technique applies a unique phase configuration across all layers in each successive time slot \cite{SIM_CE}.
Each configuration, encompassing the phase shifts of all meta-elements on all layers, transforms the channel impulse response into a distinct signal vector observed at the RF chains. 
By collecting a diverse set of these vectors over time, a sufficient number of linear equations can be constructed to solve for the channel coefficients within the current coherence time. 
This facilitates the use of subspace-based algorithms to accurately recover the high-dimensional channel despite the limited number of RF chains. The development of these estimation techniques, tailored to the specific hardware of diverse IMs, is fundamental for advancing IM-enabled ISAC systems.
\vspace{-1mm}

\subsection{IM-enabled ISAC Beamforming }
\label{sec:key}
IM-enabled ISAC can be more efficient and compact. Realizing its full potential requires beamforming strategies tailored to the unique attributes of different IM architectures.
ISAC beamforming design is usually formulated as an optimization problem focused on achieving the most effective joint communication and sensing performance metrics while considering inherent hardware limitations and power constraints. While the general objective remains consistent, the specific formulations and solution strategies vary significantly across different IM architectures. These architectural distinctions are primarily captured by the unique characteristics of their RF feeding matrix and the specific operational constraints imposed on their reconfiguration matrix, as conceptualized in our unified framework.

For RHS-enabled ISAC systems, the beamforming design balances beam pattern matching for sensing and communication with SINR threshold constraints by optimizing baseband digital precoders and the RHS reconfiguration matrix.
Decomposing the optimization into subproblems (e.g., digital precoding and reconfiguration) and employing techniques such as semi-definite relaxation and alternating optimization effectively handles the coupling between the propagation matrix and the amplitude-constrained reconfiguration matrix~\cite{RHS_1}. 

{
In contrast to the amplitude-only control of RHS, DMA-enabled ISAC beamforming must account for a Lorentzian hardware constraint that intrinsically couples amplitude and phase at each meta-element, thereby posing a distinct challenge for joint ISAC design.
A practical remedy is to handle the Lorentzian constraint via an equivalent unit‑modulus proxy variable, which recasts the design as a standard constant‑modulus optimization~\cite{DMA_1, DMA_2}.
In this formulation, the feasible set of the optimization variable becomes the product of complex unit circles, which is a Riemannian submanifold. The resulting problem can be effectively handled by manifold-optimization methods such as the Riemannian Conjugate Gradient (RCG).}

\begin{figure}[t]
    \centering
    \includegraphics[width=\linewidth]{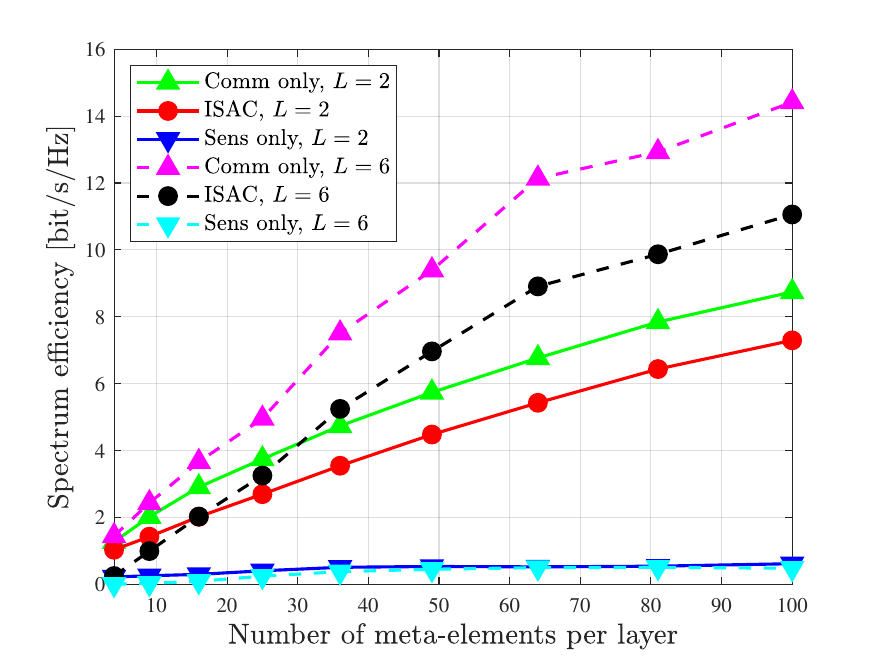}
    {
    \caption{Spectrum efficiency of communication users versus number of meta-elements.}}
    \label{fig:3}
    \vspace{-3mm}
\end{figure}

Beamforming is more complex for multi-layer architectures like SIMs due to their cascaded structure involving alternating RF feeding and reconfiguration matrices across layers.
This creates strong coupling among the phase-shift parameters of all meta-elements.
Consequently, communication and sensing signals become entangled in the wave domain as they are jointly reconfigured by multiple layers. This necessitates meticulous joint optimization to balance typically diverging ISAC objectives.
Additionally, cascaded propagation through multiple layers means the overall transmit power constraint is coupled across the entire structure. This makes it difficult to determine analog precoding matrices independently, unlike in traditional hybrid architecture antenna arrays.
Consequently, gradient-based optimization methods are commonly adopted to collectively tune the numerous coupled phase-shift coefficients in these multi-layer SIMs.
Fig.~\ref{fig:3} demonstrates how such approaches balance ISAC objectives in SIMs \cite{SIM_me}. Typically, these ISAC schemes achieve communication rates approaching those of communication-only optimization while maintaining sensing capability, with performance further enhanced by architectural improvements such as increasing the number of meta-elements per layer or, as shown, by increasing the number of metasurface layers from $L=2$ to $L=6$.
\vspace{-2mm}

\subsection{Waveform Designs for IM-enabled ISAC}
IM builds on its established capabilities in the spatial domain and introduces profound advantages by extending inherent programmability to dual-functional waveform design.
Time modulation applied to the IM elements lies at the heart of this integrated waveform design.
Time modulation is achieved by precisely controlling the temporal response of meta-elements via Time-Modulated Arrays (TMAs) in Reflective Index Surfaces (RISs) or Spatio-Temporal Coding Schemes in Space-Time-Coding Metasurfaces (STCMs). These time-varying responses effectively mix with a carrier signal. \cite{SPACE_TIME,TRIS_1}.
This interaction generates frequency harmonics alongside the fundamental frequency. The fundamental component can then be dedicated primarily to communication, while the distinct, engineered harmonics concurrently serve sensing tasks (e.g., target detection and direction of arrival estimation) with reduced mutual interference.
This mechanism is governed by specific temporal patterns applied to the IM elements and provides an elegant, hardware-efficient method for allocating frequency-domain resources directly at the waveform level. This method allows distinct functionalities to be realized from a common radiated signal.

The "coding" applied—that is, the design of these temporal patterns, such as the spatio-temporal control sequences in STCMs or the TMA parameters—is crucial. This coding not only dictates beam characteristics, but also the power levels, spatial distribution, and existence of specific harmonics. Thus, it directly sculpts the dual-function ISAC waveform. Designing these temporal and spatio-temporal codes is thus a vital aspect of IM-ISAC waveform design. It offers a software-defined approach to tailoring the spectral content and spatial properties of integrated ISAC signals to diverse operational demands and optimizing S\&C performance.

\section{Conclusions }
This paper presents a unified framework for IM-enabled ISAC systems. The framework systematically models the communication and sensing functionalities of various IM architectures, including DMA, RHS, RIS, and SIM, by analyzing their specific principles and structural features.
Building on this foundation, the article addresses key enabling technologies, including channel modeling, channel estimation, tailored beamforming, and dual-functional waveform design.
The framework unifies existing transceiver designs and provides a flexible basis for analyzing future IM variants and guiding hardware-aware algorithm development.
The framework offers valuable insights into jointly optimizing spatial, temporal, and frequency-domain resources in IM-enabled ISAC systems.
This work is expected to serve as a reference for future research and system-level integration in the rapidly evolving field of IM-enabled ISAC technologies.


\section*{Biographies}
\vspace{-15mm}
\begin{IEEEbiographynophoto}{Shunyu Li} received the B.S. degree from the School of Information and Electronics, Beijing Institute of Technology (BIT), Beijing, China, in 2021, and the M.Sc. degree from the Faculty of Science and Engineering, The University of Manchester, U.K., in 2022. He is currently working toward the Ph.D. degree with the School of Information and Electronics, BIT.
\end{IEEEbiographynophoto}
\vspace{-8mm}
\begin{IEEEbiographynophoto}{Tianqi Mao}(Member, IEEE) is currently an Associate Professor with the School of Interdisciplinary Science, Beijing Institute of Technology, Beijing, China. His research interests include modulation, waveform design and signal processing for wireless communications, integrated sensing and communication, terahertz communications, and visible light communications.
\end{IEEEbiographynophoto}
\vspace{-8mm}
\begin{IEEEbiographynophoto}{Guangyao Liu} received the B.S. degree in electronics information engineering from Dalian University of Technology, Dalian, China, in 2019. He is currently working toward the Ph.D. degree with the School of Electronic and Information Engineering, Beihang University.
\end{IEEEbiographynophoto}
\vspace{-8mm}
\begin{IEEEbiographynophoto}{Fan Zhang} received the B.S. (Hons.) degree from in 2022 the Department of Electronic Engineering, Tsinghua University, Beijing, China, where she is currently working toward the Ph.D. degree. Her main research interests include wireless communications, signal processing, waveform design in joint sensing, and communication systems.
\end{IEEEbiographynophoto}
\vspace{-8mm}
\begin{IEEEbiographynophoto}{Ruiqi (Richie) Liu}(Senior Member, IEEE) is a master researcher in the wireless and computing research institute of ZTE Corporation. He is deeply involved in specifying 5G standards through 3GPP and 6G framework through ITU, where he served as a rapporteur or chair. He is a Voting Member of the IEEE ComSoc Industry Communities Board. 
\end{IEEEbiographynophoto}

\begin{IEEEbiographynophoto}{Meng Hua}(Member, IEEE) is a Research Associate with the Department of Electrical and Electronic Engineering, Imperial College London, United Kingdom.
\end{IEEEbiographynophoto}

\begin{IEEEbiographynophoto}{Zhen Gao}(Member, IEEE) received the Ph.D. degree in communication and signal processing from Tsinghua National Laboratory for Information Science and Technology, Department of Electronic Engineering, Tsinghua University, China, in 2016. He is currently a Professor with the School of Interdisciplinary Science, Beijing Institute of Technology, Beijing.
\end{IEEEbiographynophoto}

\begin{IEEEbiographynophoto}{Qingqing Wu}(Senior Member, IEEE) is currently an Associate Professor with the Department of Electronic Engineering, Shanghai Jiao Tong University, China. He was listed as the Clarivate ESl Highly Cited Researcher in 2021 and 2022, the Most Influential Scholar Award in Al-2000 by Aminer in 2021, and World's Top 2 percent Scientist by Stanford University in 2020 and 2021.
\end{IEEEbiographynophoto}

\begin{IEEEbiographynophoto}{George K. Karagiannidis}(Fellow, IEEE)  is currently a Professor with the ECE Department of Aristotle University of Thessaloniki, Greece, and the Head and the Founder of the Wireless Communications \& Information Processing (WCIP) Group. He is one of the highly-cited authors across all areas of Electrical Engineering, recognized from Clarivate Analytics as a Highly-Cited Researcher in the ten consecutive years 2015–2024.
\end{IEEEbiographynophoto}

\end{document}